\documentclass[preprint,12pt,authoryear]{elsarticle}

\usepackage{amssymb}
\usepackage{amsmath}
\usepackage{xcolor}
\usepackage[margin=1in]{geometry}
\usepackage{lineno}

\newcommand{\bfm}[1]{\mbox{\boldmath{$#1$}}}

\journal{Icarus}

\begin{document}

\begin{frontmatter}



\title{Prediction of Apophis's deformation-driven rotational evolution during its closest encounter to the Earth in 2029} %% Article title


\author{Masatoshi Hirabayashi} %% Author name

\affiliation{organization={Georgia Institute of Technology},%Department and Organization
            addressline={}, 
            city={Atlanta},
            postcode={30332}, 
            state={Georgia},
            country={USA}}

%% Abstract
\begin{abstract}
In 2029, the near-Earth asteroid (99942) Apophis approaches the Earth within six Earth radii. This opportunity is one of the rarest natural experiments that we can use to better characterize a small body through telescopic observations and space missions. Earlier geological investigations consistently suggested that major geological processes might not occur on Apophis during this closest encounter, including surface processing and interior deformation. However, minor resurfacing may occur, depending on local geological conditions. A critical finding is that the rotational evolution occurs due to the tidal effect from the Earth. The present study offers an additional perspective on the rotational evolution, which may vary due to variations in interior properties. Namely, possible deformation processes may change the spin state variation from the rigid body state, even if deformation is not measurable. The effort in this work is to explore this issue using a simplified model, motivated by earlier studies by \cite{Hirabayashi2023} and \cite{Taylor2023}. The results show that the deformation-driven spin state change may be possible, depending on Young's modulus. If this asteroid's Young's modulus is $\sim$1 MPa or higher, the spin state only deviates a few degrees from the rigid body state over one year. However, if it is $\sim$10 kPa or less, the spin state deviation may reach a few degrees, even a few days after the closest encounter. Both telescopic observations and space missions can provide strong insights into this phenomenon.
\end{abstract}

\begin{highlights}
\item This study applies a semi-analytic approach to investigate Apophis's deformation-driven rotation during its closest encounter with Earth in 2029. 
\item If Apophis's Young's modulus is less than 10 kPa, deformation causes a rotational variation of a few degrees from the rigid body case, a few days after the closest encounter. 
\item The deformation-driven rotation mode is highly sensitive to the rotational state, needing detailed characterizations of Apophis during the closest encounter. 
\end{highlights}

%% Keywords

\end{frontmatter}



%\linenumbers
\section{Introduction}
Asteroid 99942 Apophis will closely approach Earth on April 13, 2029. This asteroid is classified as an Sq-type asteroid \citep{Binzel2007} with an equivalent diameter of $340 \pm 10$ m \citep{Brozovic2018}. The closest distance that day, according to detailed astrometric observations, is as close as six Earth radii \citep{Giorgini2008, Chesley2009, Chesley2010, Farnocchia2013}. At this distance, Earth tides may not be enough for the asteroid to experience significant geological modifications on the surface and in the interior. However, limited or moderate changes may occur depending on its configurations \citep{Yu2014, Valvano2021, kim2021surface, Ballouz2024}, which are currently unconstrained. Investigating Apophis is a rare opportunity to tackle various questions in both small body science and planetary defense \citep{Dotson2022}

One of the possible observable variations before and after the asteroid’s closest approach is its rotation state, which has been widely predicted by the scientific community \citep{Scheeres2005, Souchay2018, demartini2019using, Valvano2021, Lobanova2024, Benson2023}. The current spin is reported to be in a tumbling state \citep{pravec2014, Lee2022}, which consists of two rotational components: a spin period of $264.18 \pm 0.03$ h and a precession period of $27.3855 \pm 0.0003$ h \citep{Lee2022}. This slow, tumbling state can change after experiencing the addition of a tidal torque from the Earth during the asteroid's closest encounter. The reported tumbling mode with measurement uncertainties challenges the accurate prediction of the rotational state before and after the encounter.  

The changes in the interior during the closest encounter have been argued using various approaches. A general consensus is that this event is unlikely to result in significant interior changes \citep{demartini2019using, Hirabayashi2021}. However, this consequence may depend on the narrowness of the asteroid's possible neck; although the less severe deformation does not change \citep{Scheeres2025}. One remaining question is how the deformation mode, regardless of its small scale, contributes to the spin state \citep{Hirabayashi2023, Taylor2023}. While small, a tidal torque changes the spin state during the encounter. This spin state change also adds additional torque components. This torque causes the asteroid to deform, altering its moment of inertia (MOI). If the changed MOI remains in the post-encounter phase, angular momentum conservation leads to a change in the spin state due to variations in the MOI. Because the angular momentum is proportional to the MOI, its change also influences the angular velocity. 

The purpose of the present work is to provide a brief overview of how Apophis's deformation changes its spin state before and after the closest encounter in 2029. In our study, we simplify this problem using a toy model in which Apophis is represented by a dumbbell-shaped body, consisting of two spheres connected by a massless rod. While this model is overly simplified and thus ignores some detailed dynamics and deformation modes, it is worth investigating for the following reasons. First, the simplified model can extract key factors causing the reshaping-driven spin state change. By doing so, we can adjust key parameters to ensure the focused behaviors are consistent with sophisticated approaches. Second, using the simplified model enables a broader coverage of uncertainties. Apophis's physical and rotational states are not well constrained. The contribution of such uncertainties to its response to the closest encounter needs to be quantified. However, using sophisticated approaches often encounters significant computational burdens, which prevent investigating it. Our study aims to contribute to this effort while introducing the concept of deformation-driven spin state variation. 

We are aware that the simplification causes our results to differ from the actual behaviors, although this disadvantage may be minimal. During a gravitational encounter, an irregular shape should experience stretching along its longest axis \citep{Schaub2003}. This mode may cause the longest axis to experience the highest deformation, suggesting that focusing on the longest axis is reasonable. Nevertheless, not considering other deformation components likely underestimates the actual displacements, giving a lower bound of the rotational change. Thus, if our model suggests the possibility of the deformation-driven spin state variation, we can interpret that the actual case should exhibit a higher deviation than predicted.

The concept of deformation's contribution to the rotational state is not brand new for Apophis. A study explored photometric signature variations driven by tidal effects, considering the same concept that we explore in this study \citep{Taylor2023}. The study introduced the SAMUS simulation software, which solves the weak formulation of the partial differential Navier–Stokes equations over a finite-element mesh to determine the tidal deformation effect on a planetary object. They applied this approach to Apophis to predict a potential variation in the rotational state from the rigid body state. However, \cite{Taylor2023} stated, ``Due to the difficulties of interpreting data from this sort of lightcurve, we must develop a sophisticated non-principal axis rotation model for incorporation into SAMUS.'' The present model complements \cite{Taylor2023}, aiming to provide a better interpretation of Apophis's rotational state change during the closest encounter due to deformation. Specifically, we apply the earlier model based on continuum mechanics \citep{Hirabayashi2023} and actively incorporate Apophis's latest rotational state, including measurement uncertainties. With these new inputs, we provide a clearer understanding of how Apophis's deformation affects its rotational state, transitioning from a rigid body state. 

Our approach considers the coupling effect of both deformation and dynamical modes \citep{Hirabayashi2023}. The model can be generalized to any coupling problem; however, we apply a semi-analytic formulation \citep{Hirabayashi2023} for simplification purposes this time. In the model, we focus on characterizing the rotational response to the resulting deformation, which is controlled by stiffness and viscosity. Stiffness refers to how stiff a target is to deform. In a typical linear constitutive model, Young's modulus represents it.\footnote{In a narrow definition, stiffness is not identical to Young's modulus. Stiffness connects loading and displacement, while Young's modulus correlates between stress and strain \citep{Hughes2003}. We use stiffness as a conceptual term to describe the stiffness of Apophis, rather than applying a narrow definition.} Viscosity refers to how quickly the deformation excitation decays. The Young's modulus level is the primary quantity to be parameterized in this study. This work is organized as follows. First, we briefly define the orbital and rotational conditions during Apophis's closest approach to the Earth in 2029. Next, we introduce the simplified model employed in this study. Third, we provide simulation results and their interpretations. 

\section{Apophis's orbital and rotational state}
\label{Sec:rotation}
\begin{figure}[ht!]
    \centering
    \includegraphics[width=\linewidth]{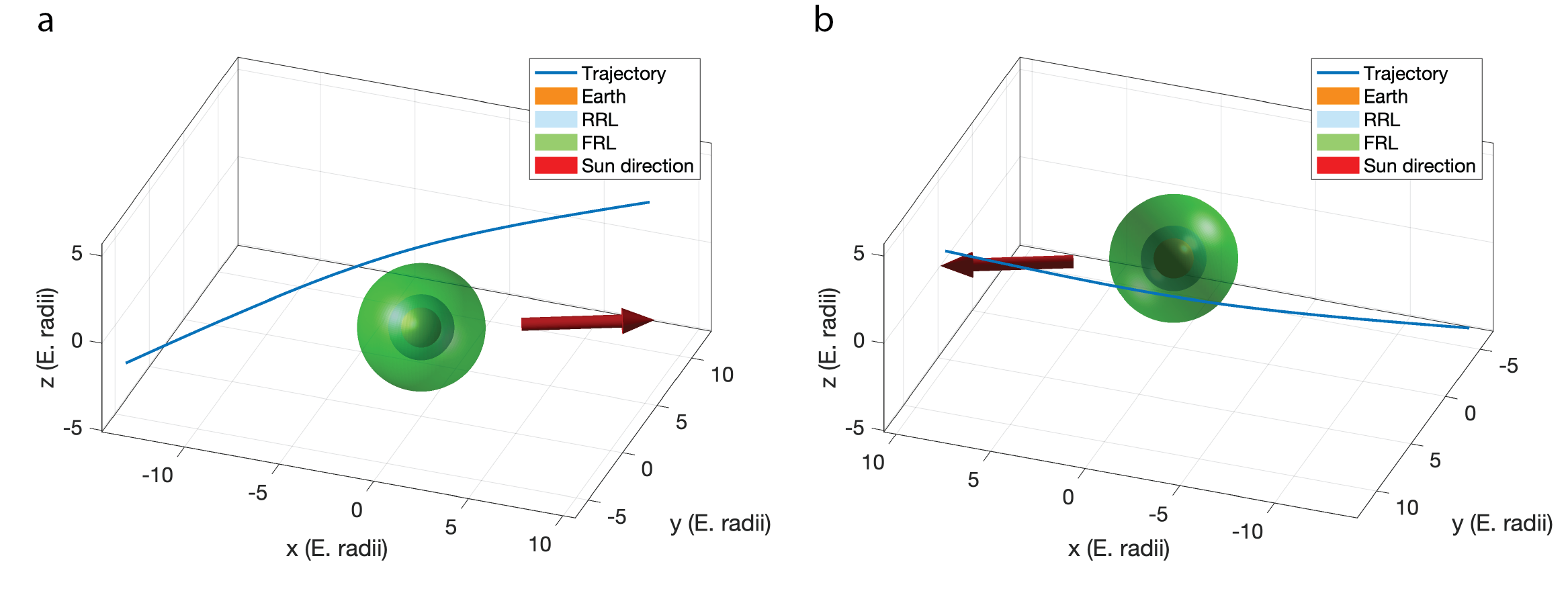}
    \caption{Apophis's orbital state during the closest approach. Panels a and b show the same condition in different views in the ecliptic J2000 coordinate frame. The asteroid passes by the Earth from left to right in Panel a and from right to left in Panel b. As of early 2025, the predicted closest approach timing is April 13, 2029, at 21:45:03.51. The blue line is the asteroid's trajectory. The axes are given in Earth radii. The red arrow points in the direction of the sun. The innermost sphere in orange is the Earth, the intermediate sphere in cyan is the rigid Roche limit (RRL), within which a rubble pile body likely disrupts due to the planet's tides, and the outermost sphere in green is the fluid Roche limit (FRL), within which aggregation processes may not create a new rubble pile.}
    \label{Fig:Apophis_Trajectory}
\end{figure}

This section reviews and obtains Apophis's orbital and rotational state during its closest encounter. The recent orbital determination suggests that Apophis approaches about six Earth radii at 21:45 on April 13, 2029 UTC. As shown in Figure \ref{Fig:Apophis_Trajectory}, the tidal effect from the Earth changes the asteroid's trajectory during the encounter. The orbital data is extracted from the Jet Propulsion Laboratory's Horizons System \citep{Giorgini1996}. Apophis approaches from Earth's southern side to the northern side; it goes from left to right in Figure \ref{Fig:Apophis_Trajectory}a and from right to left in Figure \ref{Fig:Apophis_Trajectory}b. The closest approach timing and position are given in Table \ref{Table:propA}. 

Given the closest distance, the Apophis orbit never crosses the rigid and fluid Roche limits, where the fluid Roche limit is the planet-asteroid distance, within which a rigid body tends to experience a disruption, and the fluid Roche limit defines a similar distance for a fluid object \citep{Aggarwal1974, Murray2000}. The rigid Roche limit depends on Apophis's mechanical strength and may be about 1.64 Earth radii if Apophis's bulk density is assumed to be 2,500 kg/m$^3$. This outcome comes from a typical formula, $1.26 R_E (\rho_E/\rho_A)^{1/3}$, where $R_E$ is the Earth radius, $\rho_E$ is the Earth bulk density, and $\rho_A$ is the asteroid bulk density \citep{Hesselbrock2017}. On the other hand, for the fluid Roche limit, the same bulk density suggests a fluid Roche limit of about 3.2 Earth radii. The formula for this limit is $2.456 R_E (\rho_E/\rho_A)^{1/3}$ \citep{Hesselbrock2017}. 

Recent lightcurve studies constrained Apophis's tumbling state. Table \ref{Table:rotational_state} compares the rotation state measurements at a reference epoch of 2456284.676388 JD (December 23.176388, 2012) derived by two observational groups \citep{pravec2014, Lee2022}. The latest measurements reported the asteroid's spin state consisting of a spin period of 264.18 $\pm$ 0.03 hr along the short axis and a precession period of 27.3855 $\pm$ 0.0003 hr along the angular momentum vector \citep{Lee2022}. 

\begin{figure}[ht!]
    \centering
    \includegraphics[width=\linewidth]{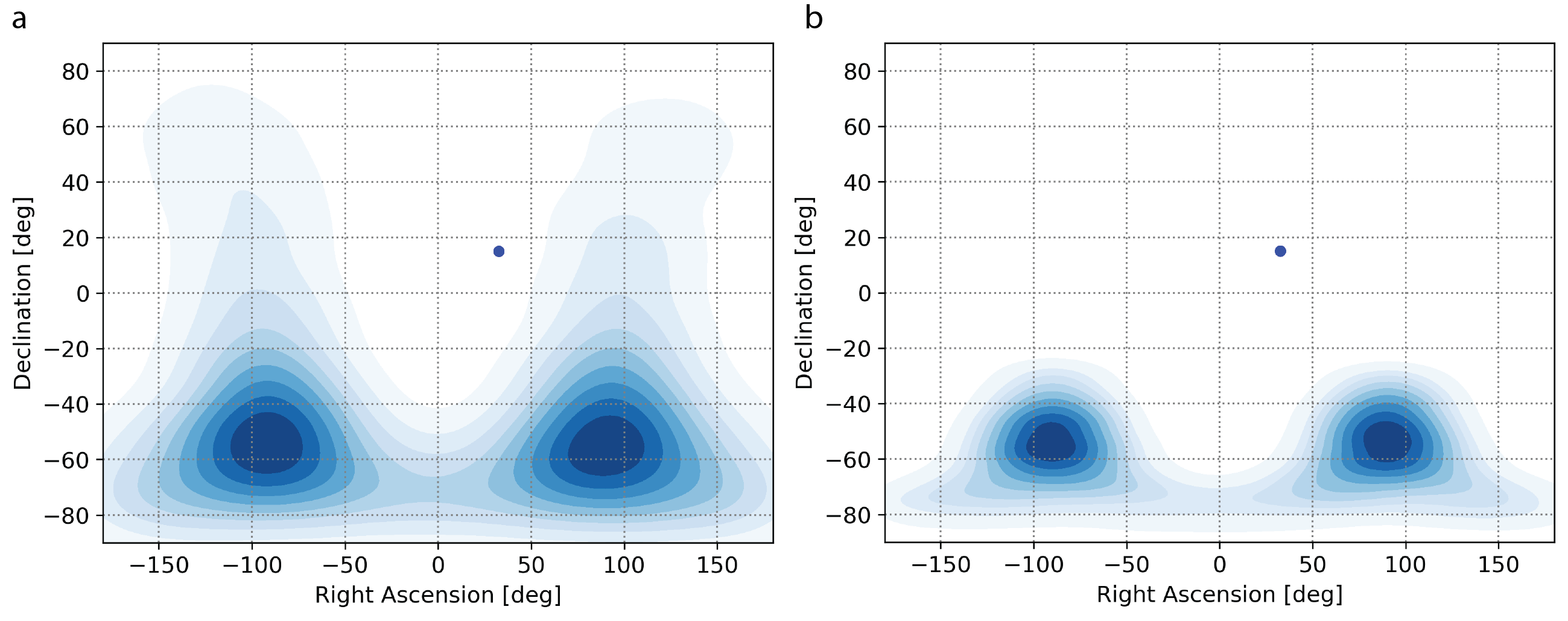}
    \caption{Spin state variations at 2029 APR 13 09:00:00 TDB, given earlier photometric observations by two groups. The $x$-axis is right ascension (RA), and the $y$-axis is declination (DEC). These quantities are provided in the ecliptic J2000 frame. Darker blue contours indicate higher probabilities. The blue dots indicate the direction of the Earth relative to Apophis (RA is 32.53$^\circ$, and DEC is 15.03$^\circ$). Panel a shows the results by \cite{pravec2014}, and Panel b indicates those by \cite{Lee2022}.}
    \label{fig:spin_comps}
\end{figure}

Using the rigid body assumption and the spin state given in Table \ref{Table:rotational_state}, we propagate the spin state from the reference epoch until 2029 APR 13 09:00:00 TDB, which is about 12.5 hr before the closest encounter (Figure \ref{fig:spin_comps}). We statistically determine the resulting spin state distribution by generating 10,000 states at the reference epoch based on measurement uncertainties. We display the two rotational predictions based on earlier measurements \citep{pravec2014, Lee2022}. The resulting spin states are likely placed in two directions just before the close encounter. One direction is a negative right ascension (RA), denoted as $\lambda$, while the other is a positive RA. This behavior results from the separatrices of the spin axis path along the intermediate principal axis, given constant angular momentum and energy \citep{Schaub2003}. The variations in energy and angular momentum levels due to measurement uncertainties allow the rotational state to cross the separatrices, resulting in two spin directions with a high probability. While both cases offer similar spin state distributions \citep{pravec2014, Lee2022}, smaller uncertainties offer smaller spin state distributions \citep{Lee2022}. The highest probability happens at $(\beta,\lambda) \sim (-55^\circ, -90^\circ)$ or $(-55^\circ, 90^\circ)$, where $\beta$ is the declination.

\begin{table}[ht!]
    \centering
    \caption {Orbital settings for simulations. The position and velocity are provided in the ecliptic J2000 coordinate frame and are relative to the Earth. TDB stands for the Barycentric Dynamical Time. } 
    \vspace{0.1in}
    \begin{small}
    \begin{tabular}{lll}
        \hline
        Properties & Value & Unites \\
        \hline \hline
        Closest distance time & 2029 APR 13 21:46:12.700 TDB & [-] \\
        Simulation start time & 2029 APR 13 9:00:00.00 TDB & [-] \\
        Simulation end time & 2029 APR 15 11:00:00.00 TDB & [-] \\
        Closest approach position  ($x$ axis) & $-1.91067934 \times 10^4$ & km \\
        Closest approach position  ($y$ axis) & $3.23019367 \times 10^4$ & km \\
        Closest approach position  ($z$ axis) & $6.03138178 \times 10^3$ & km \\
        Closest approach velocity ($x$ axis) & $6.33395774$ & km/s \\
        Closest approach velocity ($y$ axis) & $3.40222190$ & km/s \\
        Closest approach velocity ($z$ axis) & $1.84391307$ & km/s \\
        Start position ($x$ axis) & $-2.36202205 \times 10^5$ & km \\
        Start position ($y$ axis) & $-1.50660503 \times 10^5$ & km \\
        Start position ($z$ axis) & $-7.52367961 \times 10^4$ & km \\
        Start velocity ($x$ axis) & $4.33289446$ & km/s \\
        Start velocity ($y$ axis) & $3.90501392$ & km/s \\
        Start velocity ($z$ axis) & $1.69080403$ & km/s \\        
        \hline
    \end{tabular}
    \end{small}
    \label{Table:propA}
\end{table}

\begin{table}[ht!]
    \centering
    \caption{Rotational state parameters at an epoch of 2456284.676388 JD (December 23.176388, 2012), measured by \cite{pravec2014} and \cite{Lee2022}. The means directly refer to the citations, but we intentionally take a higher magnitude of uncertainty if the lower and upper ends have different separations from the mean. In this study, the right ascension (RA) is used to mean the ecliptic longitude, while the declination (DEC) is identical to the ecliptic latitude. }
    \vspace{0.1in}
    \begin{small}
    \begin{tabular}{llll}
    \hline
      Property & Notation & Value \citep{pravec2014} & Value \citep{Lee2022} \\
    \hline
    \hline
      Right ascension & $\lambda$ & $250 \pm 27$ & $278\pm10$ \\ 
      Declination & $\beta$ & $-75 \pm 14$ & $-86\pm4$ \\
       & $\phi$ & $152 \pm 173$ & $183\pm7$ \\
       & $\theta$ & $37 \pm 14$ & $49\pm15$ \\
       & $\psi$ & $14 \pm 44$ & $3\pm5$ \\
       & $P_\psi$ & $263 \pm 3$ & $264.18 \pm 0.03$ \\
       & $P_\phi$ & $27.38 \pm 0.07$ & $27.3855 \pm 0.0003$ \\
       & $I_\xi / I_\zeta$ & $0.61 \pm 0.11$ & $0.64 \pm 0.09$ \\
       & $I_\eta / I_\zeta$ & $0.965 \pm 0.015$ & $0.962 \pm 0.023$ \\
    \hline
    \end{tabular}
    \end{small}
    \label{Table:rotational_state}
\end{table}

\section{Coupling dynamics formulation} \label{sec:modeling}
The scope of this work is to provide an overview of the mechanism that causes a spin change due to deformation. The rotational state depends on the moment of inertia (MOI). If the MOI changes, it also influences the angular velocity. While detailed analysis remains necessary as future work, we focus on a simple model that can only account for the basic mechanism. The present model originates from a recent effort to generalize the coupling problem of dynamical motion and structural deformation of an irregularly shaped, self-gravitating object \citep{Hirabayashi2023}. The approach of this model is to decompose the momentum equation into three modes: translation, rotation, and deformation. The rigid body assumption is a typical approach for such decomposition, while the deformation mode needs a constitutive law as an additional constraint. The following discussion provides a summary of the technique, including the assumptions presented in this study. 

\begin{figure}
    \centering
    \includegraphics[width=0.7\linewidth]{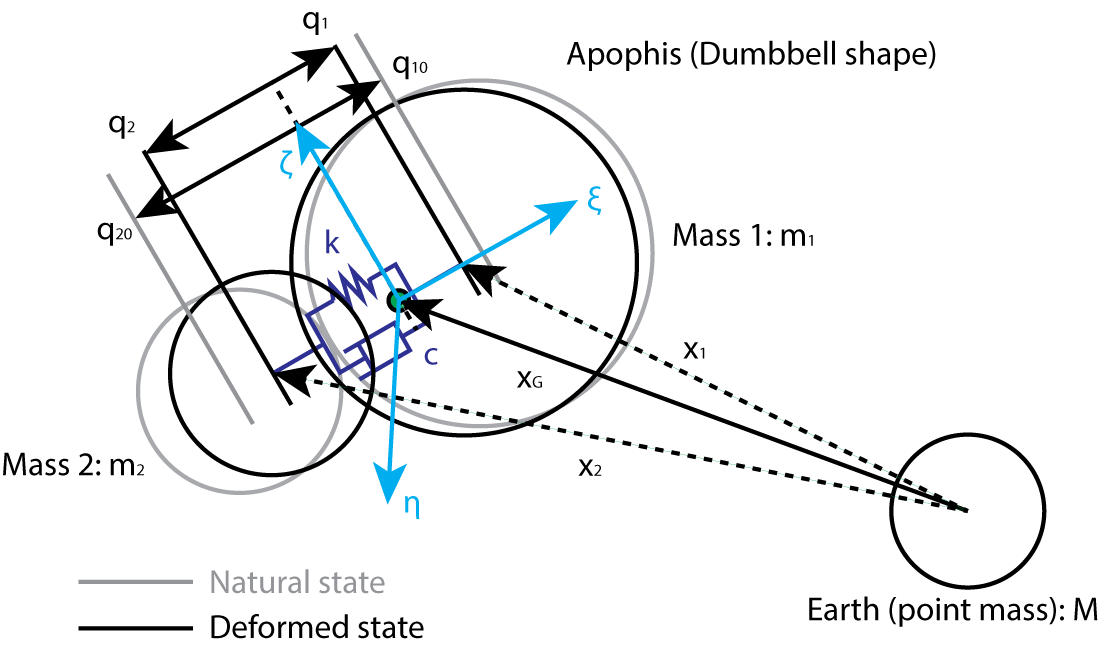}
    \caption{Schematic of the present model. }
    \label{fig:illust}
\end{figure}

Apophis is assumed to be a dumbbell-shaped body in which two spherical lobes rest on each other to exhibit the object's narrow neck. Figure \ref{fig:illust} shows the schematic of this model. The model defines the body-fixed frame aligned with the body's principal axis. The longest principal axis is the $\xi$ axis, while the intermediate and shortest axes are along the $\eta$ and $\zeta$ axes. Given the dumbbell shape's symmetry, there is no distinction between the $\eta$ and $\zeta$ axes; in other words, the MOIs along these axes are identical. However, the rotational state may have different modes along these axes. Below, we provide a brief introduction to the model formulation. 

The locations of the two lobes from the Earth, $\bfm{x}_1$ and $\bfm{x}_2$, are defined as follows:
\begin{eqnarray}
    \bfm{x}_1 &=& \bfm{x}_G + \bfm{q}_{1} = \bfm{x}_G + \bfm{q}_{10} + \bfm{u}_1 \\
    \bfm{x}_2 &=& \bfm{x}_G + \bfm{q}_{2} = \bfm{x}_G + \bfm{q}_{20} + \bfm{u}_2
\end{eqnarray}
where $\bfm{x}_G$ is the center of mass of the asteroid from the Earth, $\bfm{q}_{1}$ and $\bfm{q}_{2}$ are the lobe position vectors from the center of mass, $\bfm{q}_{10}$ and $\bfm{q}_{20}$ are the natural lobe position vectors from the center of mass, and $\bfm{u}_1$ and $\bfm{u}_2$ are the lobe displacement vectors. When the lobes' masses are given as $m_1$ and $m_2$, the forces are written as follows:
\begin{eqnarray}
    \bfm{f}_1 &=& - \frac{G M m_1}{x_1^3} \bfm{x}_1 - \frac{G m_1 m_2}{\| \bfm{q}_1 - \bfm{q}_2 \|^3} (\bfm{q}_1 - \bfm{q}_2) \\ 
    \bfm{f}_2 &=& - \frac{G M m_2}{x_2^3} \bfm{x}_2 - \frac{G m_1 m_2}{\| \bfm{q}_2 - \bfm{q}_1 \|^3} (\bfm{q}_2 - \bfm{q}_1)
\end{eqnarray}
where $M$ is the Earth's mass, and $G$ is the gravitational constant. Later, subscripts $1$ and $2$ may be replaced with $i$ to discuss further model formulation. 

Characterizing the full coupling dynamics requires accounting for three modes: translation, rotation, and deformation. For translation, the governing equation only considers the motion of the center of mass: 
\begin{eqnarray}
    m \ddot{\bfm{x}}_G = \bfm{f}_1 + \bfm{f}_2 = m \bfm{b}_t
\end{eqnarray}
where $m$ is the object's total mass, i.e., $m = m_1 + m_2$. $\bfm{b}_t$ is the total body force only contributing to translation. 

For rotation, the governing equation needs the change in the moment of inertia matrix, $\bf I$: 
\begin{eqnarray}
    ^B \dot{\bfm{\omega}} = {\bf I}^{-1} (\bfm{\tau} - ^B \dot{\bf I} \bfm{\omega} - \bfm{\omega} \times {\bf I} \bfm{\omega}) 
\end{eqnarray}
where $\bfm \tau$ is the torque acting on the target body and is given as $\bfm \tau = \bfm q_1 \times \bfm f_1 + \bfm q_2 \times \bfm f_2$, and $\bfm \omega$ is the angular velocity vector. Superscript $B$ on the left shoulder of a bold letter with a dot means a time derivative with respect to the body fixed frame, $B$. 

Finally, the deformation mode for lobe's mass $i \:(=1,2)$ characterizes the displacement evolution:
\begin{eqnarray}
   m_i ^B \ddot{\bfm{u}}_i &=& - m_i {\bf I}^{-1} (\bfm{\tau} - ^B \dot{\bf I} \bfm{\omega} - \bfm{\omega} \times {\bf I} \bfm{\omega}) \times \bfm{q}_i - 2 m_i \bfm{\omega} \times ^B \dot{\bfm{u}}_i - m_i \bfm{\omega} \times (\bfm{\omega} \times \bfm {q}_i) \nonumber \\
   &+& m_i \bfm{b}_{ri} + m_i \bfm{b}_{ui} + g_i(\bfm{\sigma}) \label{Eq:dis}
\end{eqnarray}
where $\bfm{b}_{ri}$ and $\bfm{b}_{ui}$ are the body forces contributing to the rotation and displacement of mass $i$. $g_i(\bfm \sigma)$ is the term related to the constitutive model. In this problem formulation, the spring and damper forces represent this term:
\begin{eqnarray}
    g_i (\bfm \sigma) = - k m_i \bfm u_i - c m_i \dot {\bfm u}_i 
\end{eqnarray}
where $k$ and $c$ are the spring and damper coefficients. The damping coefficient is not detailed in this work. 

The way of determining the terms on the right-hand side in Equation (\ref{Eq:dis}) is first to eliminate the body force mainly contributing to the rotational mode, i.e., $\bfm b_{ri}$, and then to determine the body force contributing to the deformation mode, i.e., $\bfm b_{ui}$, which is necessary to solve this equation. The body force acting on each body, or $\bfm b_i$, is usually given. In this problem, $\bfm b_i$ should be given as follows:
\begin{eqnarray}
    \bfm b_i = \frac{1}{m_i} \bfm f_i
\end{eqnarray}
Using this form yields the following condition:
\begin{eqnarray}
    \bfm b_i = \bfm b_{ui} + \bfm b_{ri} + \bfm b_{t} 
\end{eqnarray}
$\bfm b_{ri}$ is the term that is not yet determined but specifies the rotational mode. The term purely influencing the rotational mode is $m_i {\bf I}^{-1} \bfm{\tau} \times \bfm q_i$. Setting $\bfm b_{ri}$ as ${\bf I}^{-1} \bfm{\tau} \times \bfm q_i$ effectively eliminates the torque term. This process yields the following form: 
\begin{eqnarray}
    \bfm b_{ui} = \bfm b_i - \bfm b_{ri} - \bfm b_t = \frac{\bfm f_i}{m_i} - {\bf I}^{-1} \bfm{\tau} \times \bfm q_i - \frac{\bfm f_1 + \bfm f_2}{m}  
\end{eqnarray}

We further apply the above formulation to our problem. The principal axis MOI matrix is given as follows:
\begin{eqnarray}
    ^B{\bf I} = 
    \begin{bmatrix}
        I_\xi & 0 & 0 \\
        0 & I_\zeta & 0 \\
        0 & 0 & I_\zeta
    \end{bmatrix}
\end{eqnarray}
where $I_\xi$ is the MOI along the $\xi$ axis, and $I_\zeta$ is that along the $\zeta$ axis (and thus the $\eta$ axis). $I_\xi$ is a free parameter in this problem but does not play a critical role. On the other hand, $I_\zeta$ is given as follows:
\begin{eqnarray}
    I_\zeta = m_1 q_1^2 + m_2 q_2^2
\end{eqnarray}
where $q_1 = \| \bfm q_1\|$ and $q_2 = \| \bfm q_2\|$. $q_1$ and $q_2$ are the distances of the lobes from the center of mass, giving a condition of $m_1 q_1 = m_2 q_2$. This condition leads to $I_\zeta$ as follows:
\begin{eqnarray}
    I_\zeta = \frac{m_1}{m_2} m q_1^2
\end{eqnarray}
The time derivative of $^B \bf I$ is then given as follows:
\begin{eqnarray}
    ^B \dot{\bf I} = 
    \begin{bmatrix}
        0 & 0 & 0 \\
        0 & 2 \frac{m_1}{m_2} m q_1 \dot q_1 & 0 \\
        0 & 0 & 2 \frac{m_1}{m_2} m q_1 \dot q_1
    \end{bmatrix}
\end{eqnarray}

\section{Parameter and simulation environment settings}

\subsection{Parameter settings}
We apply the above model to investigate how Apophis responds to its Earth encounter. First, we summarize free parameters that need to be set. They include the MOI along the long axis ($I_\xi$), the lobe masses ($ m_1$, $ m_2$), the natural separation between them ($ q_{10} + q_{20}$), the spring coefficient ($k$), and the damping coefficient ($c$). $I_\xi$ only influences the rotational state along the long axis. Since no torques act on this axis, we do not discuss the rotational state along it. 

Our model needs to define $k$ and $c$ to quantify the correlation between displacement and force, thereby specifying the constitutive model. For simplicity, $c$ is set constant at $10^{-1}$ s$^{-1}$, allowing the deformed system to damp quickly in this problem. Test runs using the current model suggest that there are no significant variations in the resulting rotational states after the closest approach, unless the deformation modes contain strong oscillatory components. If the oscillatory states continue, the deformation modes continuously change the body's MOI, causing further rotational changes. 

We vary $k$ based on the conversion approach reported by \cite{demartini2019using} and \cite{DeMartini2024}, who applied the pkdgrav soft-sphere discrete element (SSDEM) code to investigate Apophis's deformation. In their SSDEM model, the spring coefficient ($k_n$) for particle interactions was set to be $\sim \pi R E$, where $R$ is the SSDEM particle radius and $E$ is Young's modulus. Introducing a rescaling factor and adjusting our notations yields our conversion function for $k$, $E$, and $m$, which is $k = 90 E / m$. The $k$ value in the present analysis varies to account for the Young's modulus range between 10 kPa and 1 MPa. We consider three $k$ values: $k = 1.51 \times 10^{-5}$ s$^{-2}$ for $E = 10$ kPa, $k = 1.51 \times 10^{-4}$ s$^{-2}$ for $E = 100$ kPa, and $k = 1.51 \times 10^{-3}$ s$^{-2}$ for $E = 1$ MPa. With this spring coefficient setting, our model outputs the displacement consistent with pdkgrav modeling \citep{demartini2019using, DeMartini2024} and finite element modeling \citep{Hirabayashi2021}, which both used the available radar shape model \citep{Brozovic2018}. Thus, while our approach utilizes the dumbbell shape as its target shape, they maintain the displacement level consistent with that of those using the radar shape model.

We fix the total mass, $m$, at $m = 6\times10^{10}$ kg, based on the radar shape model \citep{Brozovic2018} and an assumed bulk density of $2,900$ kg/m$^3$ \citep{demartini2019using}. We also keep the moment of inertia along the $\zeta$ axis, $I_\zeta$, constant at $8 \times 10^{14}$ kg m$^{-2}$. This bulk density value is slightly higher than that for other similar-sized S-type asteroids such as Asteroid Itokawa, which was reported to have a bulk density of 1,900 kg/m$^3$ \citep{Fujiwara2006}. An expected variation due to this discrepancy is approximately 35\%. The spring and damping coefficients can be rescaled using lobe masses. Because our concern is the orders of magnitude of rotational variations, such a difference is considered small. In our analysis, we consider the variation in the lobes' masses, $m_1$ and $m_2$. Importantly, however, while influencing the resulting displacements, it does not affect the spin evolution. This is a result of our simulation settings, which maintain $m$ and $I_\zeta$ as constants. We also note that our formulation of the spring and damper coefficients, which are independent of $m_1$ and $m_2$, contributes to this behavior. Details are provided in \ref{Sec:Apx1}. Table \ref{Table:prop} summarizes the properties used in our analysis.

\begin{table}[ht!]
    \centering
    \caption{Parameter settings for the present semi-analytical modeling study.}
    \vspace{0.1in}
    \begin{small}
    \begin{tabular}{llll}
    \hline
     Quantity  & Symbol & Value & Unit \\
     \hline
     \hline
     System mass & $m$ & $6 \times 10^6$ & kg \\
     Moment of inertia in the $\zeta$ axis & $I_\zeta$ & $8 \times 10^{14}$ & k m$^2$ \\ 
     Bulk density & $\rho$ & $2,900$ & kg m$^{-3}$ \\
     Damping coefficient & $c$ & $10^{-1}$ & s$^{-1}$ \\
     Spring coefficient & $k$ & $1.51\times10^{-5} - 1.51 \times 10^{-3}$ & s$^{-2}$ \\
     \hline
    \end{tabular}
    \end{small}
    \label{Table:prop}
\end{table}

\subsection{Simulation environment settings}

The primary goal of this study is to quantify the rotational angular variations of the deformation cases relative to the rigid body case by using the rotational states derived in Section \ref{Sec:rotation}. Because the proposed model accounts for deformation to track the rotational state, simply using the same initial conditions for both cases does not give consistent rotational states. This fact is critical, particularly when determining the rotational states during the closest encounter. Ideally, the rotational states of both rigid-body and deformation cases should be identical in the pre-encounter phase. In this section, we describe our approach to computing the angle deviation robustly.

We first prepare the rotational state of the deformation case without the gravitational effect of the Earth. We refer to this case as the trial case. The trial case first applies the rotational solution from Section \ref{Sec:rotation}, where we give 10,000 rotational states at 2029 APR 13 09:00:00 TDB based on observational uncertainties (Figure \ref{fig:spin_comps}). Each trial case applies a rotational state solution at 2029 APR 13 09:00:00 TDB without changing it over time to compute the equilibrium state at which deformation ceases to evolve. We run each trial case for two days to confirm that the deformation evolution has stabilized. The trial case offers the initial settings for the following two simulation cases: the rigid-body case and the deformation case. The rigid-body case only considers the asteroid's orbital and rotational states. The deformation case accounts for the deformation mode in addition to the other states. Both cases update their initial settings using the trial case and propagate the asteroid's state until 2029 APR 15 11:00:00.00 TDB (Table \ref{Table:propA}).

We compute the rigid body case by using the axis lengths and MOIs from the trial case. This case considers the gravitational effect from the Earth, which becomes maximum 12.5 hours after the simulation begins and continues for an additional 37.5 hours, resulting in a total simulation time of 48 hours. In other words, the closest approach is at 12.5 hours in the simulation. We save a solution for this case, including the angle and angular velocity evolution over each principal axis. We select the 12.5-hour scale before the closest encounter to ensure that the initial state is not affected by the Earth's gravity. If we consider a shorter timescale, such as three hours before the closest encounter, the gravitational effect from the Earth is not negligible; thus, the angular deviation starts immediately after the start of the simulation.

In the deformation case, the nominal parameters are the same as those in the rigid-body case, but they account for the displacement vectors during the settlement phase. This case also accounts for the gravitational effect of the Earth, under the same conditions as the rigid body case. If deformation is negligible, this case is ideally identical to or practically similar to the rigid body case. Otherwise, we anticipate angular variations in the rotational state from the rigid body case. The deformation-driven spin state deviation is calculated by subtracting the rigid-body case from the deformation case for rotation along each principal axis.

\section{Results}

We investigate multiple cases that change the spring coefficient while keeping the damping coefficient constant. Again, we investigate three cases of spring coefficients. The first case is $1.51 \times 10^{-3}$ s$^{-2}$, equivalent to a Young's Modulus of 1 MPa. The second case is $1.51 \times 10^{-4}$ s$^{-2}$, equivalent to a Young's Modulus of 100 kPa. The final case is a spring coefficient of $1.51 \times 10^{-5}$ s$^{-2}$, equivalent to a Young's modulus of 10 kPa. Throughout the simulations, the damping coefficient is fixed at $10^{-1}$ s$^{-1}$. Our analysis parameterizes $m_1$ and $m_2$ while keeping $m$ and $I_\zeta$ constant. We report the displacement results for three $m_1/m_2$ cases: 0.43, 0.67, and 1.0. However, we only show the spin state solutions for the case of $m_1/m_2 = 1$, as they do not change with $m_1/m_2$. Later, we also explore 10,000 rotational cases from Section \ref{Sec:rotation} to see how the rotational state changes the deformation-driven spin variation.

\begin{figure}[ht!]
    \centering
    \includegraphics[width=\linewidth]{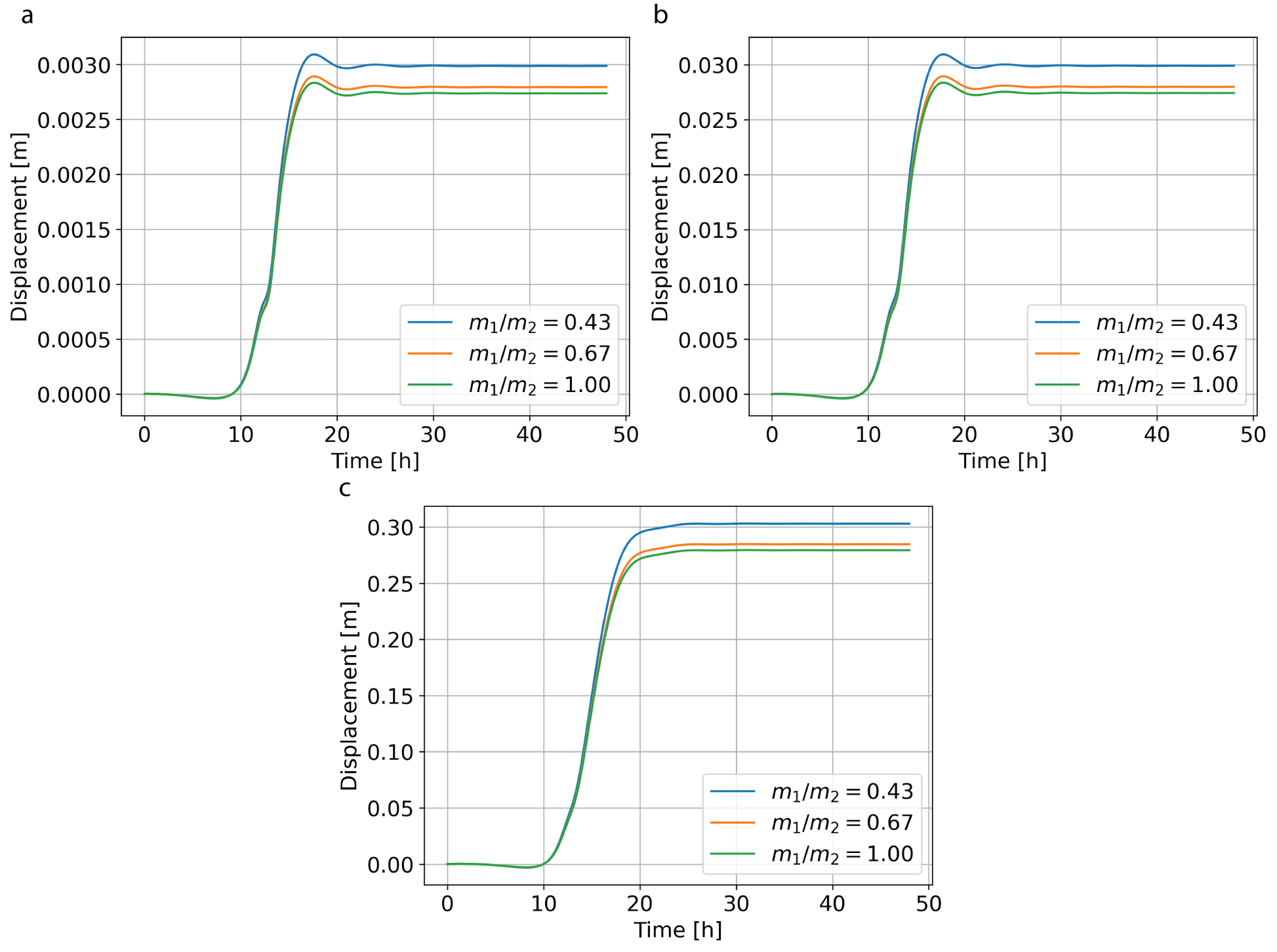}
    \caption{Time evolution of displacement along the long axis, which is the $\xi$ axis, in meters. The $x$ axis is the time in hours, while the $y$ axis gives the displacement in meters. Panel $a$ displays the displacement evolution for a spring coefficient of $1.51 \times 10^{-3}$ s$^{-2}$ ($E = 1$ MPa). Panel $b$ illustrates that evolution for a spring coefficient of $1.51 \times 10^{-4}$ s$^{-2}$ ($E = 100$ kPa). Panel $c$ depicts the same for a spring coefficient of $1.51 \times 10^{-5}$ s$^{-2}$ ($E = 10$ kPa). The damping coefficient is $10^{-1}$ s$^{-1}$ for all cases. Each panel also gives the displacement evolution for three $m_1/m_2$ cases. The blue lines are for $m_1/m_2 = 0.43$, the orange lines are for $m_1/m_2 = 0.67$, and the green lines are for $m_1/m_2 = 1.0$.}
    \label{fig:disp}
\end{figure}

Figure \ref{fig:disp} shows the displacement evolution along the $\xi$ axis. Again, all simulations start 12.5 hours before the closest approach, and the displacement is provided based on the rigid body case as the reference state. About the first 10 hours, the displacement is almost zero. However, after that time, it rapidly increases so that the $\xi$ axis becomes longer. The displacement does not return to the zero state, i.e., the original asteroid length. Instead, the displacement reaches a new settlement; after this point, the displacement does not change. The primary reason for this behavior is that rotational change generates different centrifugal forces, resulting in a new, distinct equilibrium state within the structure. The magnitude of displacement is almost proportional to the spring coefficient, which is understandable. There is an overshoot around 17.5 hours after the simulation starts. This behavior results from the relationship between the spin and damping coefficients controlling the oscillatory mode. The $m_1/m_2$ variation also changes the final displacement for each spring coefficient case. However, its contribution is limited compared to the rotational effect.

\begin{figure}[ht!]
    \centering
    \includegraphics[width=\linewidth]{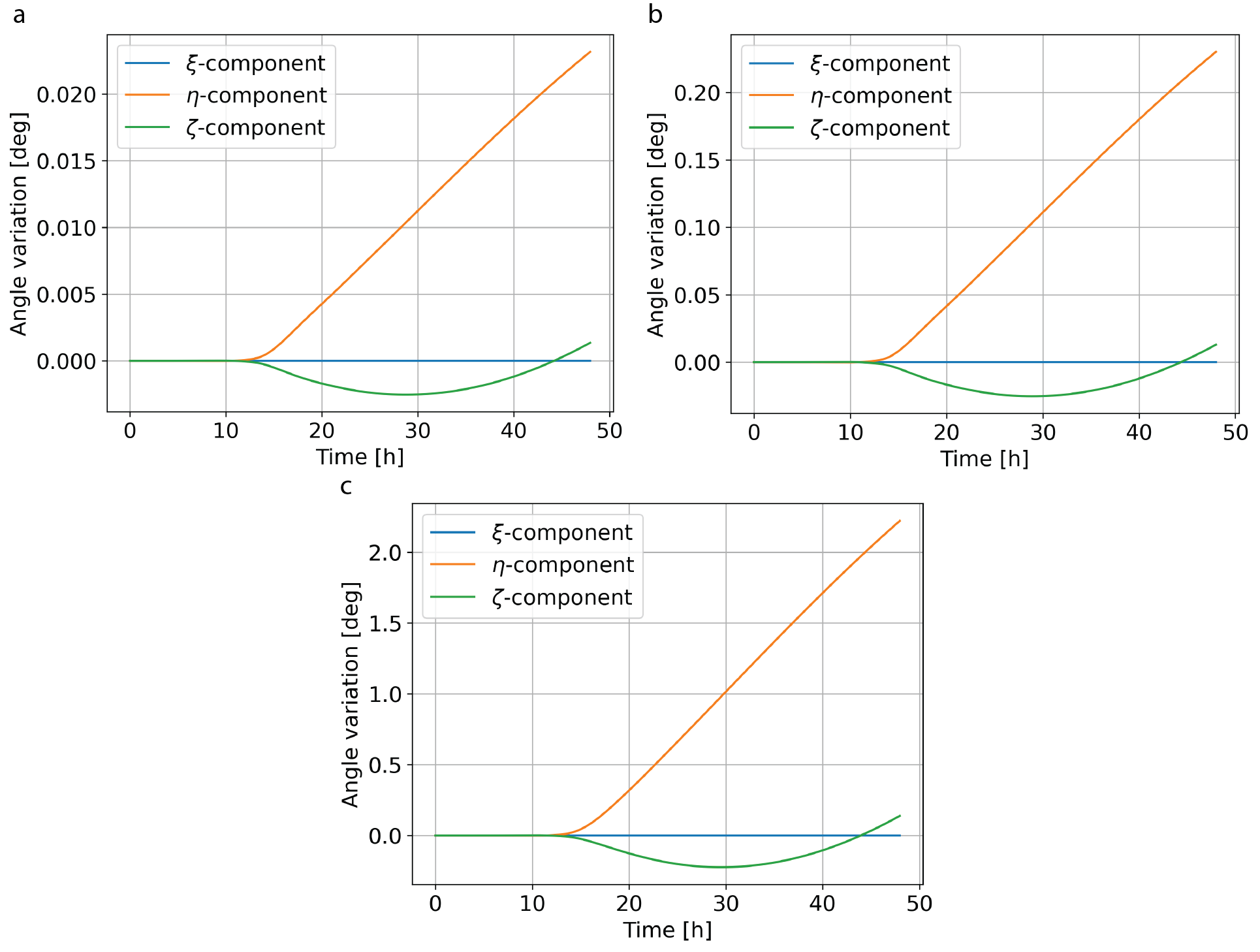}
    \caption{Angular variations with time along three principal axes. The blue lines show the angular variations along the $\xi$ component. The yellow lines display the angular variations along the $\eta$ component. Lastly, the green lines are the angular variations along the $\zeta$ component. The $x$ axis shows the time in hours while the $y$ axis describes the angular variation in degrees. The time settings are the same as in Figure \ref{fig:disp}. Panel a shows the case with a spring coefficient of $1.51 \times 10^{-3}$ s$^{-2}$ ($E = 1$ MPa). Panel b displays the case with a spring coefficient of $1.51 \times 10^{-4}$ s$^{-2}$ ($E = 100$ kPa). Panel c depicts the case with a spring coefficient of $1.51 \times 10^{-5}$ s$^{-2}$ ($E = 10$ kPa). The damping coefficient is $10^{-1}$ s$^{-1}$ for all cases. All the panels show the cases for $m_1/m_2 = 1.0$.}
    \label{fig:ang}
\end{figure}

Figure \ref{fig:ang}, on the other hand, shows the time evolution of the angular variations along the principal axis. Each panel shows the results from a given spring coefficient under a constant damping coefficient. As stated, we only show the $m_1/m_2$ cases as the results do not change. The rotation along the $\xi$ component is unaffected, given the fact that our model considers a dumbbell shape; therefore, there is zero effect on rotation along this axis. The $\eta$ component continuously increases; on the other hand, the $\zeta$ component behaves differently. This rotational variation behavior depends on the spin state at 2029 APR 13 09:00:00 TDB. The magnitude of the angular variation is proportional to the spring coefficient. As the spring coefficient increases, the angular variations decrease. An important finding from this test is that when we consider a spring coefficient of $1.51 \times 10^{-5}$ s$^{-2}$, equivalent to a Young's modulus of 10 kPa, the angular variations reach a few degrees within a few days after the closest encounter. This finding is critical because, depending on Apophis's stiffness, future telescopic observations and exploration missions may be able to characterize its rotational variations from the rigid body state.

We further characterize statistical trends of the deformation-driven spin state variation. We investigate angular variations when the spring coefficient is $1.51 \times 10^{-5}$ s$^{-2}$ and the damping coefficient is kept constant at $10^{-1}$ s$^{-1}$. The target of this statistical investigation is to account for the rotational uncertainties provided by observational measurements. We utilize the spin state derived from observational measurements in Figure \ref{fig:spin_comps}, comprising a total of 10,000 cases, to examine how the rotational uncertainties impact the rotational variations. Similar to Figure \ref{fig:ang}, the following discussions only show the cases for $m_1/m_2 = 1.0$.

\begin{figure}
    \centering
    \includegraphics[width=\linewidth]{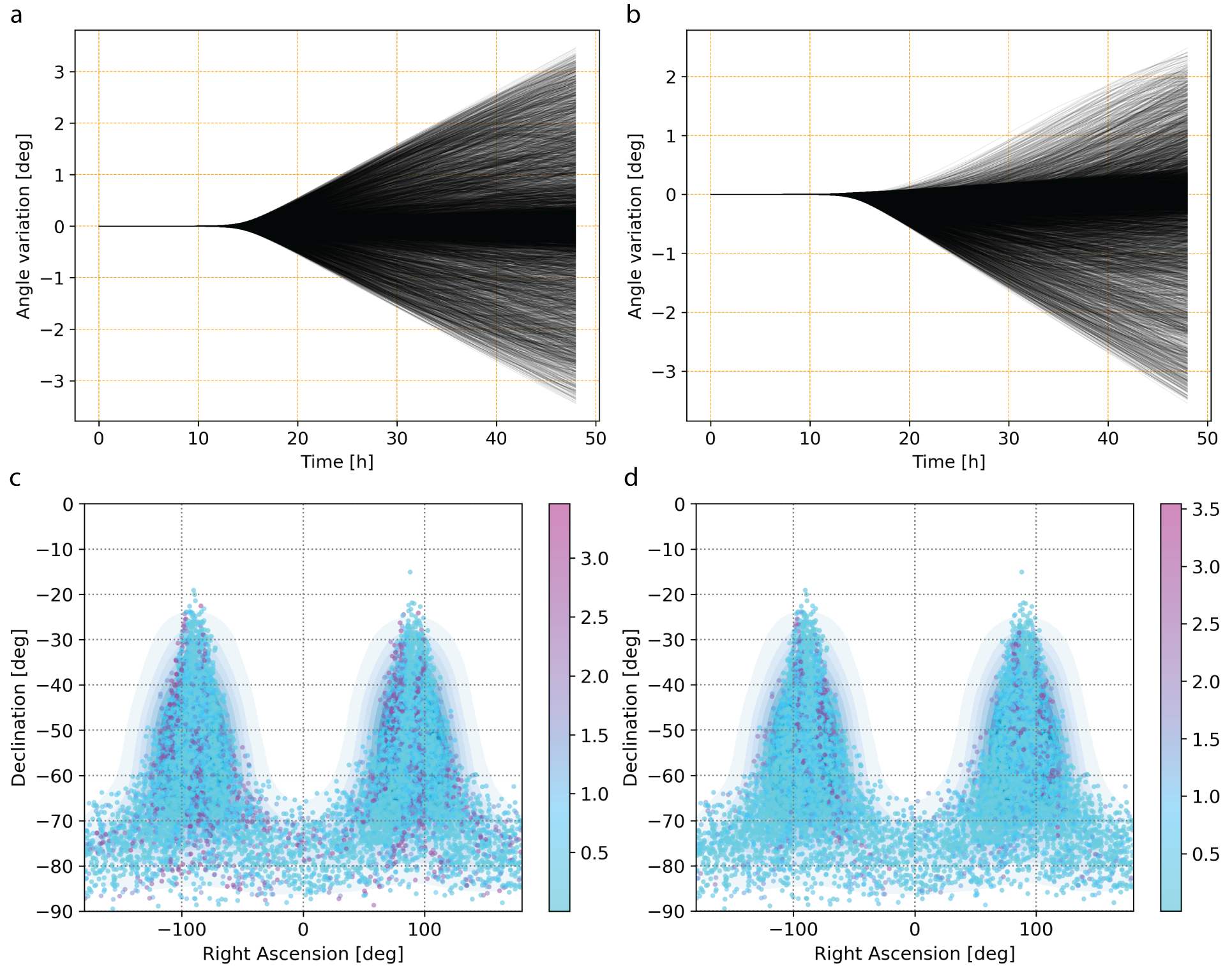}
    \caption{Statistical analysis using the rotational distributions from Figure \ref{fig:spin_comps}. Panels a and b show the angle variations for 10,000 cases using the derived rotational states at the beginning of the simulation time. Each black line gives the angular variation over time for a given rotational state within measurement uncertainties. Panel a shows the angular variation along the $\eta$ component, while Panel b describes that along the $\zeta$ component. Panel c and d map the maximum angle variations onto the rotational distribution map in Figure \ref{fig:spin_comps}. The color bars define the maximum angle variation in degrees. Panel c gives the results for the $\eta$ component, which are identical to Panel a. On the other hand, Panel d describes the results for the $\zeta$ component coming from Panel b. All the panels show the cases for $m_1/m_2 = 1.0$.}
    \label{fig:stat}
\end{figure}

Figure \ref{fig:stat} shows the statistical trends of the angular variations along the $\eta$ and $\zeta$ axes. The time evolutions of the angular variations are almost symmetrical to the zero states, though some biases appear in the $\zeta$ component. Lower angular variations may be possible, as there exist dense distributions (Figures \ref{fig:stat}a and \ref{fig:stat}b). Figures \ref{fig:stat}c and \ref{fig:stat}d map the maximum angular variations onto the spin state distributions derived from measurement uncertainties (Figure \ref{fig:spin_comps}). Trends are almost similar, but there are no strong correlations between the spin state's highly dense areas and high angular variations. In other words, high angular variations are not concentrated around the dense region of the spin states. This finding suggests that determining the deformation mode and resulting rotational variation may be challenging until detailed observations are made available.

\section{Discussions}

In this study, we investigated Apophis's potential deformation mode, which causes rotational variations from the rigid body case. Rather than detailing the mechanism, we focused on introducing a basic concept of it. With this, we applied a simple model in which Apophis's shape was treated as a dumbbell shape. This model is simple yet reasonable, given that earlier studies employed a similar model for a contact binary and provided meaningful insights \citep{Hirabayashi2010, Hirabayashi2023}. The results inferred that depending on the stiffness of the structure, Apophis might exhibit interesting phenomena in which it deforms and has a different rotational state than the rigid body case. The key finding was that the displacement might not have to be large or measurable by spacecraft, but small, non-negligible deformation processes could alter the moment of inertia and change the angular velocity vector at a given time, resulting in angular variations.

According to our model, if Apophis's Young's modulus is around 10 kPa, the angular variations from the rigid body state may reach a few degrees a few days after the closest encounter, depending on the rotational state. If its Young's modulus is 100 kPa, the angular deviation may be a few tenths of degrees a few days after the closest encounter. A higher Young's modulus makes Apophis stiffer, so the angular variations become negligible. However, taking longer measurements may enable the detection of such deviations. The derived 10 kPa Young's modulus is in general smaller than that for typical granular media on the Earth, which may be $\sim$MPa or higher \citep{Lambe2008}. However, if Apophis is a rubble pile, which is consistent with observational interpretations \citep{Muller2014}, how its loose structure in a microgravity environment controls stiffness is unknown. This finding is critical because many sub-kilometer asteroids are known to be rubble piles \citep{Fujiwara2006, Watanabe2019, Barnouin2019, Levison2024} and are common in the near-Earth region. Detailing Apophis, more specifically the proposed phenomenon on this asteroid, can provide stronger insights into the interiors of such sub-kilometer asteroids.

The concept underlying this phenomenon is not based on resurfacing. Recent analysis suggested that the expected spin change may lead to non-negligible resurfacing processes, allowing the surface conditions to settle into a different state after the encounter \citep{Ballouz2024}. While their idea may sound similar to our concept, it differs slightly because processes are not necessarily irreversible (such as resurfacing or internal failure). Both reversible and irreversible deformation can change the spin state. Once the spin state changes after the closest encounter, deformation should occur to establish a new equilibrium state, thereby furthering the spin state change. Therefore, if such a deformation is not negligible, the resulting angular variations should appear immediately. 

This work explores the same problem as provided by \cite{Taylor2023}. While our approach differs essentially from theirs, the conclusions are consistent: if Apophis meets the stiffness condition, deformation-driven spin state variations may be observable. Our study explicitly accounted for the elastic model, whereas \cite{Taylor2023} applied the SAMUS simulation software to solve the incompressible Navier-Stokes equation, which takes into account irreversible deformation, a concept similar to that of \cite{Ballouz2024}. Our intention to stick to the elastic mode stems from earlier studies \citep{demartini2019using, DeMartini2024, Hirabayashi2021}, which all indicated that irreversible deformation would be negligible. If inelasticity dominates, the timescale of lightcurve detection may be longer, as found by \cite{Ballouz2024}. Our study suggests that the detection timescale depends on Young's modulus. If Young's modulus is 1 MPa, the spin state may only deviate a few degrees even after one year. However, if it is about 10 kPa, the deviation may reach 90$^\circ$ within two months, depending on the rotational state at the closest encounter. Furthermore, in contrast to \cite{Taylor2023}, who primarily investigated simpler principal axis rotation modes, we examined how the asteroid's complex tumbling state influences deformation-driven spin state variations. The key finding is that the tumbling mode plays a critical role in changing them, suggesting that pre-encounter spin state measurements may be greatly helpful.

While our analysis complements earlier studies, we add a note regarding an open question for future investigations. The parameter choice by \cite{Taylor2023} resulted in a change of up to 1 m in Apophis's axis, while our selection of the parameter settings, mainly the spring coefficient, provided an axis change of up to 30 cm. This 30-cm axis change was consistent with predictions by analysis using the pkdgrav SSDEM code, particularly when Apophis consisted of a randomly packed structure \citep{demartini2019using, DeMartini2024}, and finite element modeling \citep{Hirabayashi2021}. This displacement consistency came from our setting of the $k$ quantity, allowing the model to represent the displacement level for the radar shape model \citep{Brozovic2018}, rather than that for an extreme contact binary shape. Using our model yields an axis change of 1 m if Young's modulus is $\sim$1 kPa. While all these studies offer comparable values, whether the proposed range of a low Young's modulus is appropriate for a sub-kilometer-sized rubble pile asteroid needs further investigation. Additionally, the magnitude of the axis change that is reasonable for observational measurements needs to be quantified in future studies \citep{Taylor2023}. 

The major issue of identifying this phenomenon is that it is necessary to decompose Apophis's spin state into the deformation and rigid body spin states. A typical approach to determining the spin state involves continuously monitoring it via telescopic observations. However, just doing so cannot constrain these two spin elements. This is because such measurements do not determine the contribution of deformation to the spin state. One way to tackle this issue is for future missions to detail the shape and interior during the closest encounter, and for telescopic observations to monitor the spin state continuously for a longer timescale \citep{Reddy2022, Taylor2023}. For space missions, determining the moment of inertia at least before or after the closest encounter is crucial. Gravity field characterizations, seismometers, and radar tomography can offer strong insights into such measurements. If there is a strong spin deviation from the rigid body case, such a signature can be identified as angular variations. Again, if Young's modulus is around 10 kPa, the anticipated angular variations may increase up to 10s of degrees in months and 100s of degrees in a year.

A better understanding of Apophis' rotational angular variations contributes to both scientific and planetary defense investigations. Space explorations to sub-kilometer asteroids have been unsuccessful in offering direct insight into their interiors, except for recent gravity measurements by OSIRIS-REx for the near-Earth asteroid (101955) Bennu \citep{scheeres2020heterogeneous}. If successful, the resulting angular variations can provide direct insights into Apophis's bulk strength property, which is key information about the interior structure of sub-kilometer asteroids. Determining an asteroid's mechanical strength is also a critical contribution to planetary defense. The strength parameter is one of the key quantities for planetary defense to assess potential deflection technologies to be used \citep{NEOWARP2024}. Using this approach and detailing Apophis's possible deformation-driven rotational variations from its rigid-body state is beneficial to science and planetary defense investigations planned for future Apophis missions \citep{dellagiustina2023osiris, Kuppers2024, Morelli2024, Andrade2025, Daca2025}.

\section{Conclusion}
This study examined the deformation-driven rotational variation that may occur in Apophis during its 2029 encounter with the Earth. We applied a semi-analytical model that decomposes the momentum equation into three components: translation, rotation, and deformation. The model was simplified so that the target body was assumed to be a dumbbell shape, consisting of two equally massive, spherical lobes attached by a massless rod. The tidal effect from the Earth was generated using a point mass in the model. Although this model overlooked several components, it was successful in characterizing the key components of deformation-driven rotation. Our parametric approach, which varied the asteroid's stiffness while maintaining its viscosity, revealed that stiffness could control the rotational angular variations. The results showed that the deformation-driven spin state change strongly depended on Young's modulus. If this asteroid has a Young's modulus of $\sim$1 MPa or higher, the spin state only deviates a few degrees from the rigid body state over one year. However, if it is $\sim$10 kPa or less, the spin state deviation might reach 90$^\circ$ a few months after the closest encounter. We also conducted a statistical analysis, considering the rotational state uncertainties, and found that the rotational state would be a critical factor in altering the rotational variations. Unless the detailed determination of the rotational state is available, determining the deformation level and thus the resulting deformation-driven rotation state accurately may be challenging. Performing telescopic observations and space exploration missions can detail this phenomenon.

\appendix

\section{Analytical validation of the present model}
\label{Sec:Apx1}

Although the present model is overly simplified, we utilize it with numerical quantities to describe a relatively complex phenomenon, particularly in accounting for the gravitational interaction between Apophis and the Earth. This challenge prevents perfect validation, but further simplification can offer some analytical insights. We consider that a dumbbell shape constantly spins at a spin rate of $\omega$ along the $\zeta$ direction in the pre-encounter condition. Then, this body flies by the Earth to acquire an additional angular momentum along the $\zeta$ axis, $\Delta h$, thereby changing its configuration. This section uses the quantities defined above; however, if not provided, we define additional properties anew. 

We first formulate the change in angular momentum. The initial angular momentum along the $\zeta$ axis is written as $h = I_\zeta \omega$. Then, the addition of $\Delta h$ to $h$ yields the configuration change as follows.
\begin{eqnarray}
    (I_\zeta + \Delta I_\zeta) (\omega + \Delta \omega) = h + \Delta h
\end{eqnarray}
For simplicity, we consider $\omega$ to be positive, while the defined variations for $h$, $I_\zeta$, and $\omega$, i.e., $\Delta h$, $\Delta I_\zeta$, and $\Delta \omega$, respectively, can take both positive and negative values. Using a first-order approximation yields the following condition for $\Delta \omega$, $\Delta h$, and $\Delta I_\zeta$:
\begin{eqnarray}
    \Delta \omega \sim \frac{1}{I_\zeta} (\Delta h - \omega \Delta I_\zeta) \label{Eq:angMom}
\end{eqnarray}
This form characterizes how $\Delta \omega$ varies with $\Delta h$ and $\Delta I_\zeta$, given $I_\zeta$ and $\omega$ for the pre-encounter phase. It is reasonable as $\Delta \omega$ increases with the addition of a positive $\Delta h$ while decreasing with a higher moment of inertia, i.e., $\Delta I_\zeta > 0$. 

We next consider the force balance on $m_1$ affected by the spring, rotation, and gravity. The force balance prior to the closest approach may be given as follows:
\begin{eqnarray}
   m_1 k u_1 + m_1 \omega^2 q_1 = \frac{G m_1 m_2^3}{q_1^2 m^2} \label{Eq:balance1}
\end{eqnarray}
For simplicity, all the values above are considered positive, assuming that gravity is dominant. Simplifying Equation (\ref{Eq:balance1}) yields:
\begin{eqnarray}
   k u_1 q_1^2+ \omega^2 q_1^3 = \frac{G m_2^3}{m^2} \label{Eq:balance2}
\end{eqnarray}

We now consider $\omega$, $q_1$, and $u_1$ change to $\omega + \Delta \omega$, $q_1 + \Delta q_1$, and $u_1 + \Delta u_1$, respectively, where $\Delta \omega$, $\Delta q_1$, and $\Delta u_1$ can take both positive and negative values. From our simulation settings, because $\Delta q_1 = \Delta u_1$, we use $\Delta u_1$ for $\Delta q_1$ below. Using a first-order approximation gives the following form:
\begin{eqnarray}
   \Delta \omega \sim - \frac{3 \omega^2 + k}{2 \omega} \frac{\Delta u_1}{q_1} - \frac{k}{\omega} \frac{u_1}{q_1} \frac{\Delta u_1}{q_1}
   \label{Eq:balance2}
\end{eqnarray}
This equation highlights a key feature of how $\Delta \omega$ varies with $\Delta u_1$. When $\omega$, $k$, $u_1$, and $q_1$ are provided for the pre-encounter phase, a positive $\Delta \omega_1$ offers a negative $\Delta u_1$, meaning that the displacement becomes smaller under the gravity-dominant condition. 

Equations (\ref{Eq:angMom}) and (\ref{Eq:balance2}) provide two key features in our formulation: First, if $I_\zeta$ and $m$ are set constant, $\Delta \omega$ only depends on $\omega$ and $k$, not $m_1$ and $m_2$. In Equation (\ref{Eq:angMom}), $\Delta h$ is proportional to $\Delta I_\zeta$, given gravity gradient \citep{Schaub2003}. Thus, if $\Delta I_\zeta$ is constant, then $\Delta \omega$ is constant. Because our settings guarantee the same $I_\zeta$, the system should experience the same rotational state evolution, giving the same $\Delta I_\zeta$ regardless of $m_1$ and $m_2$, as long as they maintain $m$ constant. Furthermore, Equation (\ref{Eq:balance2}) has $\Delta u_1 / q_1$, where $\Delta I_\zeta / I_\zeta = 2 \Delta u_1 / q_1$, guaranteeing $\Delta \omega$ to be constant. This indicates that $\Delta \omega$ is a function of $\omega$ and $k$. Second, with this constant, $\Delta u_1$ changes with $m_1$ and $m_2$. Equation (\ref{Eq:balance2}) shows that while $\Delta \omega$ stays constant from the earlier discussion, $\Delta u_1$ can change due to the existence of $q_1$, which is dependent on $m_1$ and $m_2$. 

Although it is a crude validation, we incorporate the spin state evolution and $k = 1.51 \times 10^{-5}$ s$^{-2}$ used for the case presented in Figures \ref{fig:disp} and \ref{fig:ang} into Equation (\ref{Eq:angMom}). The $\Delta \omega$ value from the simulation is $-1.39 \times 10^{-4}$ s$^{-1}$, while our prediction using Equation (\ref{Eq:angMom}), multiplying a factor of $\sqrt{2}$ to account for the axis symmetry along the $\eta$ and $\zeta$ axes, gives $-1.17 \times 10^{-4}$ s$^{-1}$. We confirm that both quantities do not change with $m_1$ and $m_2$, given constant $m$ and $I_\zeta$. This convenient behavior also stems from our setting of the spring and damper coefficients, which are independent of $m_1$ and $m_2$.

\newpage

%\bibliography{manuscript}
%\bibliographystyle{elsarticle-harv}

\end{document}